**Improving homology-directed repair by small molecule agents for genetic engineering in unconventional yeast? - Learning from the engineering of mammalian systems.**

Min Lu[1] and Sonja Billerbeck[1]*

[1]Molecular Microbiology, Groningen Biomolecular Sciences and Biotechnology Institute, University of Groningen, Groningen, The Netherlands

*Correspondence: s.k.billerbeck@rug.nl

**Abstract**

The ability to precisely edit genomes by deleting or adding genetic information enables the study of biological functions and the building of efficient cell factories. In many unconventional yeasts – such as promising new hosts for cell factory design but also human pathogenic yeasts and food spoilers – this progress has been limited by the fact that most yeasts favor non-homologous end joining (NHEJ) over homologous recombination (HR) as DNA repair mechanism, impairing genetic access to these hosts. In mammalian cells, small molecules that either inhibit proteins involved in NHEJ, enhance protein function in HR, or molecules that arrest the cell cycle in HR-dominant phases are regarded as promising agents for the simple and transient increase of HR-mediated genome editing without the need for *a priori* host engineering. Only a few of these chemicals have been applied to the engineering of yeast although the targeted proteins are mostly conserved; making chemical agents a yet underexplored area in enhancing yeast engineering. Here, we consolidate knowledge of available small molecules that have been used to improve HR efficiency in mammalian cells and the few ones that have been used in yeast. We include available high throughput (HTP) -compatible NHEJ/HR quantification assays that could be used to screen for and isolate yeast-specific inhibitors.

**Graphical Abstract**

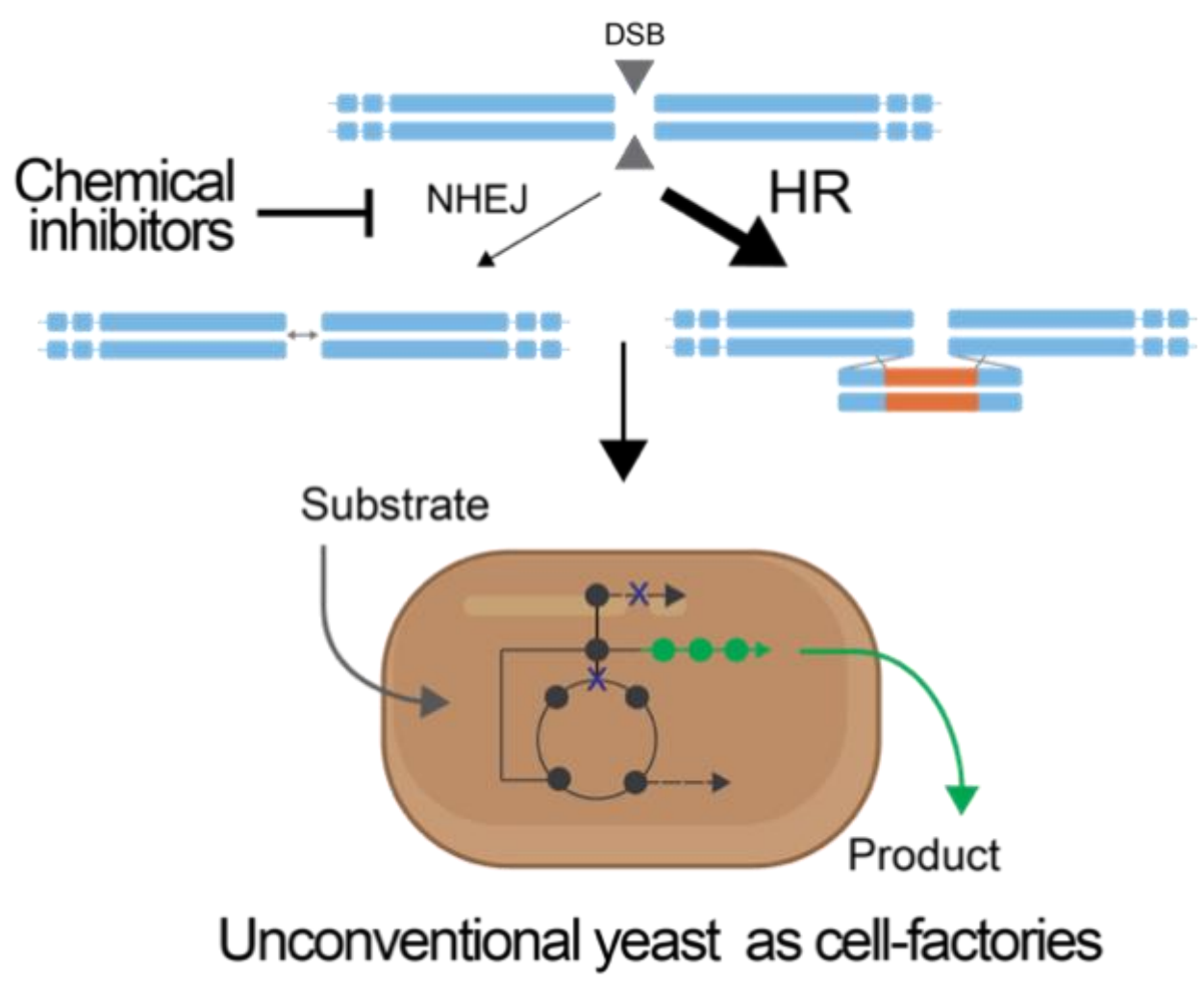

## 1. Introduction

DNA double-strand breaks (DSB) induced by radiation, toxic chemicals, or endogenous processes, are the most serious DNA damage resulting in genome instability and cell death (Cannan, et al., 2016). To survive this injury, two major repair pathways are rapidly activated – non-homologous ending joining (NHEJ) or homologous recombination (HR) **(Figure 1A).** NHEJ is active during the whole cell cycle and is regarded as an error-prone pathway, causing the random insertion or deletion of bases and thus potentially the disruption of open reading frames. HR is a more precise repair mechanism where broken ends are ligated with the help of homologous donor DNA. HR works more efficiently in the S and G2 phases of the cell cycle where homologous sister chromatids are available (Saha, et al., 2017). Even though there is competition between NHEJ and HR repair (Azhagiri, et al., 2021), NHEJ always acts kinetically faster and dominates in mammalian cells and most yeast species.

Two exceptions are the yeasts *Saccharomyces cerevisiae* and *Schizosaccharomyces pombe* which – at least the well-established laboratory strains – favor HR during all stages of the cell cycle. The fact that *S. cerevisiae* primarily employs HR for DSB repair has turned this yeast into a living machine for DNA assembly up to genome size (Gibson, et al., 2008) and has facilitated the development of simple, marker-less, and multiplexable genome editing tools that enabled refactoring of entire pathways and implementation of synthetic metabolism (Malci, et al., 2020).

Although *S. cerevisiae* has been extensively engineered for commercial manufacturing (Kavscek, et al., 2015; Nielsen, 2019), various of the other 1,500 known yeast species gained attention as potential cell factories. This is because these unconventional yeasts show a natural capacity to produce desired commodity chemicals, a broad capacity to use various substrates and their often more robust process characteristics. Recent reviews (Cai, et al., 2019; Park, et al., 2022) highlight examples such as the oleaginous yeast *Yarrowia lipolytica* for lipid production, *Kluyveromyces marxianus* for its thermotolerance and broad substrate utilization, *Scheffersomyces stipitis* for its natural capacity for xylose utilization and production of aromatic compounds, and the methylotrophic yeasts *Hansenula polymorpha* and *Pichia pastoris* for their extremely efficient heterologous protein secretion and glycosylation ability.

Further – besides the field of biotechnology – several unconventional yeasts are pathogens to humans or food spoilers. Examples include *Candida albicans*, *C. tropicalis*, *C. auris*, *C. glabrata*, *Zygosaccchoromyces rouxii*, and *balii or Brettanomyces bruxiliensis.*

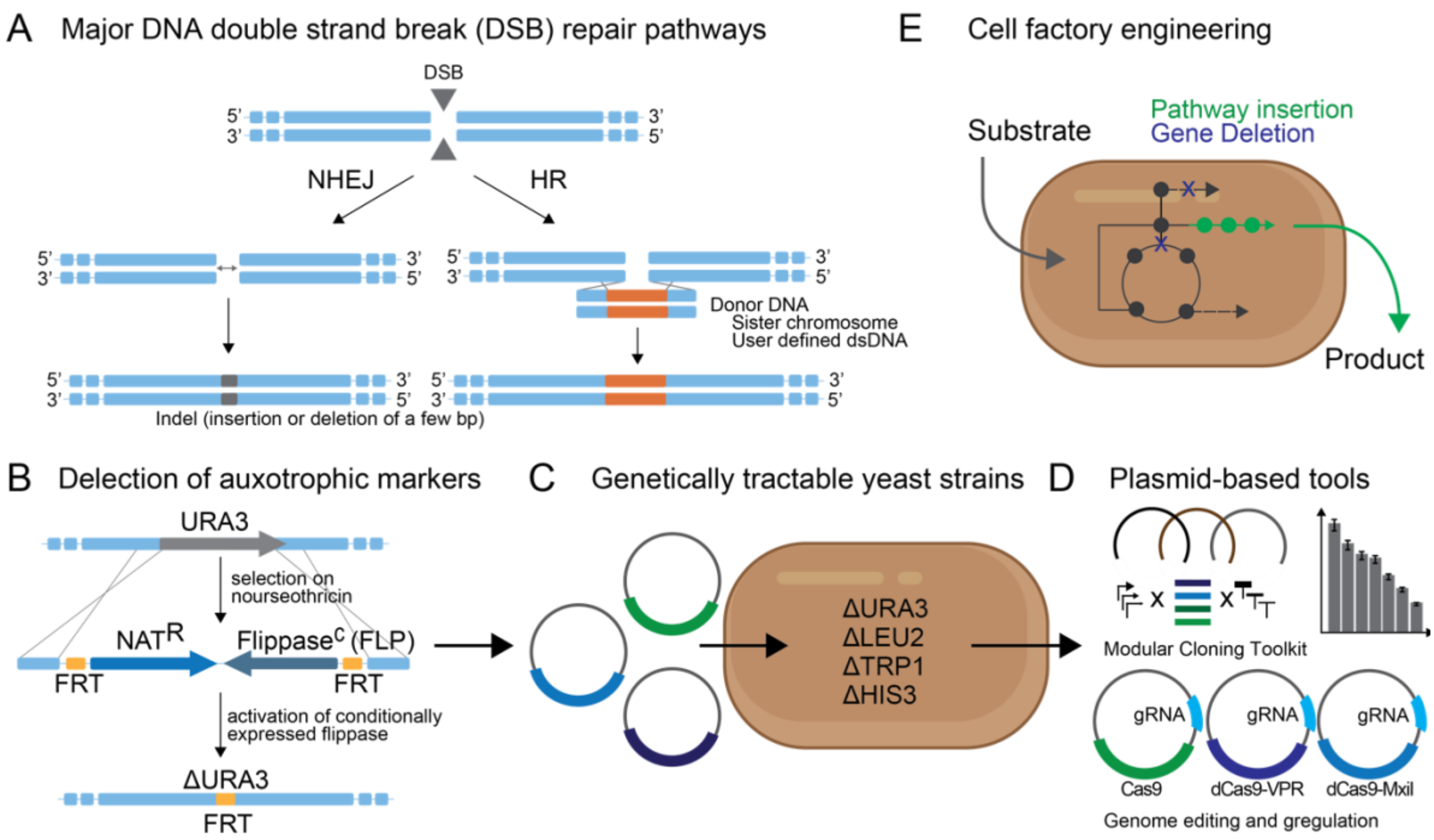

**Figure 1. HR-mediated repair of a DSB by a user-designed donor DNA is a powerful tool to gain genetic access to unconventional yeast and turn them into cell factories. A.** Simple overview of the two major repair pathways that are activated after a DSB. A DSB can be induced anywhere in the genome by radiation or chemicals or it can be induced at a specific locus by genome editing tools such as CRISPR/Cas, TALENs, or the I-SceI endonuclease. Nonhomologous end-joining (NHEJ) leads to error-prone ligation of the broken DNA ends that can cause indels (insertions or deletions of a few base pairs) at the repaired site (see (Dudášová, et al., 2004) for a detailed overview in yeast). Homologous recombination (HR) uses a homologous donor DNA to repair the cut. This donor DNA can be derived from homologous sister chromatids or by creating a user-defined fragment that encodes extensions (homologous arms) that are homolog to a given locus. **B.** HR-mediated repair can be harnessed to delete auxotrophic markers such as the URA3 gene (encoding for the enzyme Orotidine 5-phosphate decarboxylase, which catalysis one step in the biosynthesis of pyrimidine ribonucleotides). The donor DNA usually encodes for an antibiotic selection marker $NAT^R$ (giving resistance to nourseothricin) and a conditionally expressed flippase recombinase (FLP) that is used to eventually remove the $NAT^R$ gene via recombination with reciprocal FRT sites and recycle it for the deletion of other auxotrophic markers (examples include (Schwarzmuller, et al., 2014; Moreno-Beltran, et al., 2021). **C and D:** One or multiple deleted auxotrophic markers allow to transform a yeast with one or multiple plasmids that encode the deleted gene for selection on minimal media lacking the appropriate metabolite (Brachmann, et al., 1998; Billerbeck, et al., 2023). Auxotrophic markers are commonly used in yeast engineering as only a few antibiotics markers are available. Plasmids allow the building of modular cloning toolkits and the use of genome editing and genome regulation tools. **E.** Modular gene expression, deletion, and insertion tools are essential for cell-factory engineering via the insertion of expression-balanced multi-gene pathways and deletion of interfering metabolic reactions.

Infections caused by opportunistic pathogenic yeast can be detrimental especially to immunocompromised individuals, resulting in 80.000-2.3 million cases every year (Gabaldon, et al., 2019). In addition, antifungal resistance has been rising and multi-resistant strains of *C. glabrata* and *C. auris* have emerged (Fisher, et al., 2016; Gabaldon, et al., 2019).

Genetic tractability of unconventional yeast is thus not only fundamental for metabolic engineering and gaining access to new cell factories for a sustainable bio-economy but also for answering basic biological questions about pathogenicity, host-microbe interactions, for promoting understanding of antifungal mechanisms of action and ultimately to developing novel antifungal drugs.

Homology-directed repair is an essential tool that allows access to the genetics of a given host at all stages; from the use of plasmids to genome editing **(Figure 1B-C)** and its absence in a cell can make simple engineering difficult: For example, HR is a powerful (almost essential) tool for making strains accessible for plasmid transformation by deleting commonly used auxotrophic markers The use of plasmids is then the first step in developing well-characterized modular cloning toolkits(Lee, et al., 2015; Billerbeck, et al., 2023) which have proven useful for cell factory engineering (Besada-Lombana, et al., 2018) and for using plasmid-encoded genome editing or regulatory tools such as CRISPR/Cas9 and CRISPRi/a. Beyond plasmids, HR-mediated DSB repair via a donor DNA is an essential requirement for precise genome editing such as insertion of pathways or deletion of enzymatic functions **(Figure 1D)**: fundamental steps in cell-factory engineering. Of note, multiplexing of genome editing in *S. cerevisiae* has been explicitly easy because relatively short homology arms (~40 bps) are sufficient to mediate HR, and those can be encoded on primers allowing a user to cheaply generate any donor DNA by a one-step PCR. In several other yeasts, HR can only be detected when longer homology arms are provided (~500-1000 bps) (Schwarzmuller, et al., 2014; Ji, et al., 2020; Moreno-Beltran, et al., 2021), which requires more laborious fusion or extension PCRs.

Due to the essential nature of HR for genome engineering and the compromising fact that HR is outcompeted by NHEJ in most non-*S. cerevisiae* yeasts and also in mammalian cells, several efforts have focused on enhancing repair bias towards HR by either impairing the NHEJ machinery, by enhancing the HR machinery or by harnessing the cell-cycle S/G2 phase-specific HR dominance. Most strategies are genetic: many studies in yeast and mammalian cells have demonstrated that deletion of the Ku70/80 complex – an important and conserved component of the NHEJ machinery – leads to a much higher HR-mediated gene knock-in (KI) efficiency (Kooistra, et al., 2004; Ninomiya, et al., 2004; Nayak, et al., 2006; Arras, et al., 2016; Nambiar, et al., 2019). Further, overexpression of the *S. cerevisiae* ScRAD51 or ScRAD52 – two key components of the very efficient HR machinery in *S. cerevisiae* – enhanced repair via HR in mammalian cells (Vispe, et al., 1998; Shao, et al., 2017) and yeast (*Y. lipolytica*) (Ji, et al., 2020). In addition, a Geminin-based tag has been developed that restricts Cas9 expression and thus DSB induction to the S/G2 cell cycle phase and this tag has been successfully employed in human and yeast cells with enhanced HR efficiency (Ploessl, et al., 2022). Although these genetic methods improve the low HR efficiency, they require *a-priori* access to the host's genetics which is often not available. Also, the deletion of the Ku70/80 affects cell growth, and even telomere instability (Sui, et al., 2020; Ploessl, et al., 2022).

An alternative is the use of small molecules that transiently bias repair to HR. The effects of these small molecules mirror the effects of the genetic methods: they directly inhibit NHEJ-related proteins, enhance the activity of HR-mediating proteins, or synchronize and arrest the cell cycle in the S/G2 phase. Small molecules can be purchased at an affordable cost **(Supplementary Table 1)**, and employed without the need for laborious plasmid constructions or genetic changes, further, their action is transient and reversible such that wild-type genotypes for functional studies or cell factory engineering can be maintained.

Small molecules that inhibit NHEJ have been widely developed for cancer therapy (Hengel, et al., 2017) and the same molecules have subsequently been successfully employed in mammalian cell engineering (Chen, et al., 2022; Shams, et al., 2022). Given the fact that most proteins involved in NHEJ and HR are conserved (Dudášová, et al., 2004; Li, et al., 2008), it seems worthwhile exploring if these small molecules can be successfully applied in yeast engineering, and/or how chemical agents optimized for yeast can be identified. Mammalian studies already indicate that these molecules can cross the cytosolic membrane and gain access to the nucleus. Given their small molecular weight **(Supplementary Figure 1),** they likely also diffuse through the yeast cell wall which is reported to be quite permeable to large molecules including proteins in actively growing cells (Denobel, et al., 1991).

In this review, we introduce small molecules that have been successfully applied in mammalian cells and feature the few studies that have tested these molecules in yeast. We summarize HR/NHEJ quantification assays that could be repurposed to find new or modified molecules that are specifically yeast-specific molecules.

As a note, the reported efficiencies of the various chemicals in human cells are difficult to compare given their dependence on cell line, used concentration, varying length of homology arms and loci (Shams, et al., 2022), and the fact that no standardized assays for measuring “precise engineering” exist, but assays have rather been geared towards answering a specific question or meeting a specific goal of a given study.

## MAIN TEXT

### 2. Inhibitors of NHEJ.

In eukaryotes repair of a DSB by NHEJ starts with the Ku70/Ku80 heterodimer – the DNA-dependent protein kinase regulatory factor – binding the two loose DNA ends, acting as a structural linker that protects them from degradation (Dudášová, et al., 2004) **(Figure 2).** In humans, the Ku70/Ku80 heterodimer then recruits the catalytic subunit DNA-PKcs, which is thought to induce conformational changes that allow end-processing enzymes to access the DNA ends. *S. cerevisiae* does not encode a homolog of DNA-PKcs. Instead, a complex consisting of Mre11, Rad50, and Xrs2 (short MRX) is thought to perform this function, a similar complex also participates in mammalian NHEJ (MRE11/RAD50/NBS1; short MRN). In both species, the process of NHEJ is completed by the DNA ligase IV-XRCC4 complex which ligates the broken ends of the DNA, depending on the overhangs of the cut, this can be error prone or clean (Dudášová, et al., 2004). Most small molecule inhibitors developed for mammalian cells target one of the four major NHEJ components: the K70/80 heterodimer (directly or indirectly), the DNA-PK catalytic subunit, Mre11 (parts of the MRN/MRX complex) or Ligase IV, three of which are shared amongst yeast and mammalian NHEJ. Here we summarize those molecules that have been used in mammalian engineering (Chen, et al., 2022; Shams, et al., 2022)– some successful, some showing no effect, some with varying results in different cell lines – and yeast **(Table 1 and Figure 2).**

**2.1. Inhibitors of the Ku70/80 heterodimer showed compound and cell-line dependent enhancement of HR-mediated knock-in (KI) rates in human cells but they have not been tested in yeast.** Mammalian and yeast Ku70/80 are conserved and its deletion has shown significant enhancement of HR-mediated genome editing in many species, including yeast (Kooistra, et al., 2004; Ninomiya, et al., 2004; Nayak, et al., 2006; Arras, et al., 2016). A computational small-molecule screen against a potential binding pocket within the human Ku70/80 heterodimer yielded the small molecule STL127705 which showed strong inhibition of Ku70/80 binding to DNA and inhibition of the Ku-dependent activation of the DNA-PKcs kinase *in vitro* and in human cell lines, as well as higher cellular susceptible to radiation, indicating a clear potential to diminish DSB repair by NHEJ (Weterings, et al., 2016; Guo, et al., 2022). STL127705 has not been tested in genome editing studies. However, its derivative STL127685 was tested for CRISPR-based genome editing and exhibited no significant effect on targeted nucleotide substitution (TNS) efficiency in hiPSC lines (Riesenberg, et al., 2018), but has not been tested in yeast. To the best of our knowledge only STL127705 would be commercially available for these studies **(Supplementary Table 1)**. An independent computational structure-guided small-molecule screen targeting the DNA-binding activity of the Ku70/Ku80 heterodimer (so-called Ku DNA binding inhibitors, short Ku-DBi's) yielded compounds 68, 149, and 322 which inhibited the Ku-DNA interaction *in vitro* (Gavande, et al., 2020). A further study showed that multiple Ku-DBi 's sensitized non-small cell lung cancer (NSCLC) cells to DSB-inducing agents (Mendoza-Munoz, et al., 2023). The utility of various Ku-DBis was also tested via CRISPR-based genome editing increasing the KI efficiency in mouse and human cells (Gavande, et al., 2020). Neither of the Ku-DBi's has been tested in yeast. Given most Ku70/80 inhibitors have been developed and optimized by computational design, and given structural information of the human and *S. cerevisiae* Ku70/80 heterodimer (PDBs 1JEQ, 1JEY, 5Y58) are available, docking studies using ALFA-fold predicted Ku70/80 structures from

various non-conventional yeast could potentially yield insights into their binding affinity or guide the design of better fitting inhibitors.

**2.2. InsP6-depletion via calmodulin inhibition enhanced HR-mediated KI efficiencies in the two tested yeast species in a species-dependent manner:** Inositol hexakisphosphate (InsP6) is a potent cofactor of the Ku70/Ku80 activity in NHEJ. InsP6 binds to Ku70/Ku80 and induces a conformational change required for Ku70/80 mobility and migration into the nucleus where it can participate in DNA end-joining activity (Byrum, et al., 2004). The biosynthesis of the small molecule InsP6 requires calmodulin. The compound W7 – short for N-(6-aminohexyl)-5-chloro-1-naphthalenesulfonamideis – and Chlorpromazine are both calmodulin inhibitors that were shown to deplete InsP6 in Hela cells and to inhibit Ku mobility (Byrum, et al., 2004). Thus W7 and Chlorpromazine were considered potent agents for augmenting HR efficiency (Byrum, et al., 2004), but to the best of our knowledge, these compounds have not been tested in human genome editing. The high level of conservation between human and yeast calmodulin (Davis, et al., 1989), inspired the testing of their effect on genome engineering in two yeast species: in *Cryptococcus neoformans* W7 increased HR-based KI rates by 3- to 5-fold and chlorpromazine showed a 2.5 to 4-fold enhancement both depending on the locus and the concentration (Arras, et al., 2016). Both molecules subsequently allowed the authors to achieve routine homology-directed gene deletions with >50% success rate across other loci (Arras, et al., 2016). In one strain of *Metschnikowia pulcherrima*, W7 showed a very moderate 2% increase in HR-mediated integration rate in a concentration-dependent manner, chlorpromazine was not tested (Moreno-Beltran, et al., 2021).

**2.3. Ligase IV inhibition showed variable effects in human cell lines and has been tested in one yeast species with no effect.** Ligase IV connects broken DNA ends in the final step of NHEJ. A computational docking screen using a homology model of Ligase IV yielded the potential inhibitor molecules Scr7 and several derivates that subsequently showed to block NHEJ *in vitro* and to decrease cancer progression *in vivo* (Srivastava, et al., 2012). Although controversial results exist about its activity and specificity (Greco, et al., 2016), Scr7 was successfully tested to increase CRISPR-based KI rates up to 19-fold in various human/mammalian cell lines and mouse zygotes (Maruyama, et al., 2015). Independent studies showed 1.7 and 3-fold increase in KI rates in different cancer cell lines (Hu, et al., 2018; Anuchina, et al., 2023) and up to 15% increase in zebrafish embryos (Zhang, et al., 2018). while other studies report no effect on gene insertion efficiencies (Song, et al., 2016; Gerlach, et al., 2018). Notably, all studies use different cell lines, loci, and protocols (Ray, et al., 2020). The efficacy of Scr7 was also determined in the pathogenic yeast *C. neoformans* (in the same assay as W7 and Chlorpromazine), but no significant increase in KI rate was observed (Arras, et al., 2016). A derivative of Scr7, Scr130, with increased ligase IV specificity, and a 20-fold increased inhibitory coefficient was

developed, that effectively inhibited NHEJ in cancer cells (Ray, et al., 2020) but that has not been tested yet in genome editing studies.

**2.4. Inhibition of Mre11 with Mirin reduces NHEJ in mammalian cells and enhanced HR in the one tested yeast species:** Mre11 is part of the MRE11/RAD50/NBS1 (MRN) complex in humans and the Mre11/Rad50/Xrs2 (MRX) complex in yeast cells. Human and yeast MRE11 are homologs. The MRN/MRX complex is multifunctional and possesses 3′ to 5′ dsDNA and ssDNA endonuclease, ssDNA exonuclease, and hairpin cleavage activities which have essential roles in both repair pathways, NHEJ and HR **(Figure 2)**. A chemical docking screen using the Mre11 3D structure from the eubacterium *Thermotoga maritima* as a guide identified the molecule Mirin as an inhibitor of the MRN complex (Dupre, et al., 2008). Studies with Mirin analogs, designed to specifically inhibit either its exonuclease or endonuclease activity show that Mre11 plays a critical role in the cellular choice over a repair pathway (Shibata, et al., 2014). A CRISPR-assisted DSB-repair screen showed that Mirin reduced NHEJ and its alternative pathways at various cut sites (Taheri-Ghahfarokhi, et al., 2018), but to the best of our knowledge, it has not been tested in genome homology-directed engineering studies in human cells. However, it was tested in the yeast *C. neoformans*and showed a > 30% increase in KI rates (Arras, et al., 2016).

**2.5. Several inhibitors that target proteins involved in cellular decision-making over NHEJ vs. HR have been successfully employed in human cells but not tested in yeast.** The corresponding protein targets ATR, ATM, Chk1, and 53BP1 all have homologs in yeast. ATR (yeast homolog: Mecl) and ATM (yeast homolog: Tel1) regulate checkpoint kinases Chk1 and Chk2 (yeast homologs: Chk1 and Rad53), which are vital anticipators in the DNA damage response and there is an assumption that ATM or ATR inhibition can facilitate biasing decision making towards HR (Jin, et al., 2019; Wang, et al., 2021). In a small-molecule screen, Ma *et al*. identify the molecules VE-822 (inhibitor of ATR) and AZD7762 (inhibitor of Chk1) that increased CRISPR-Cpf1-mediated KI rates by 5.9- and 2.7-fold, respectively, in human pluripotent stem cells (Ma, et al., 2018). Trichostatin A (inhibitor of ATM) increased the targeted nucleotide substitution efficiency by 1.5- to 2.2-fold in hiPSC lines when using a Cas9n approach (Cas9n is a nickase that introduces a single strand break, when used with two reciprocal gRNA it in their sum nicks both strands) but not when using regular Cas9 for DSB induction (Riesenberg, et al., 2018).

Another control point in decision-making over repair pathways is protein neddylation. In neddylation, the small peptide NEDD8 gets conjugated to a protein to target it for proteasomal degradation (Maghames, et al., 2018). Within DSB repair regulation, it was shown that Neddylation inhibits CtIP-mediated resection and biases repair towards NHEJ while inhibition of Neddylation changes the normal repair profile towards an increase in HR. NEDD8-activating enzyme (NAE) is an essential regulatory component of the NEDD8 conjugation pathway and the small molecule MLN4924 was identified as a potent and selective inhibitor of NAE (Soucy, et al., 2009). MLN4942 very minimally enhanced precise genome editing by 1.1 to 1.3-fold, but it was shown to work additive in combination with other small molecules when mixed (Riesenberg, et al., 2018) **(and Point 5).**

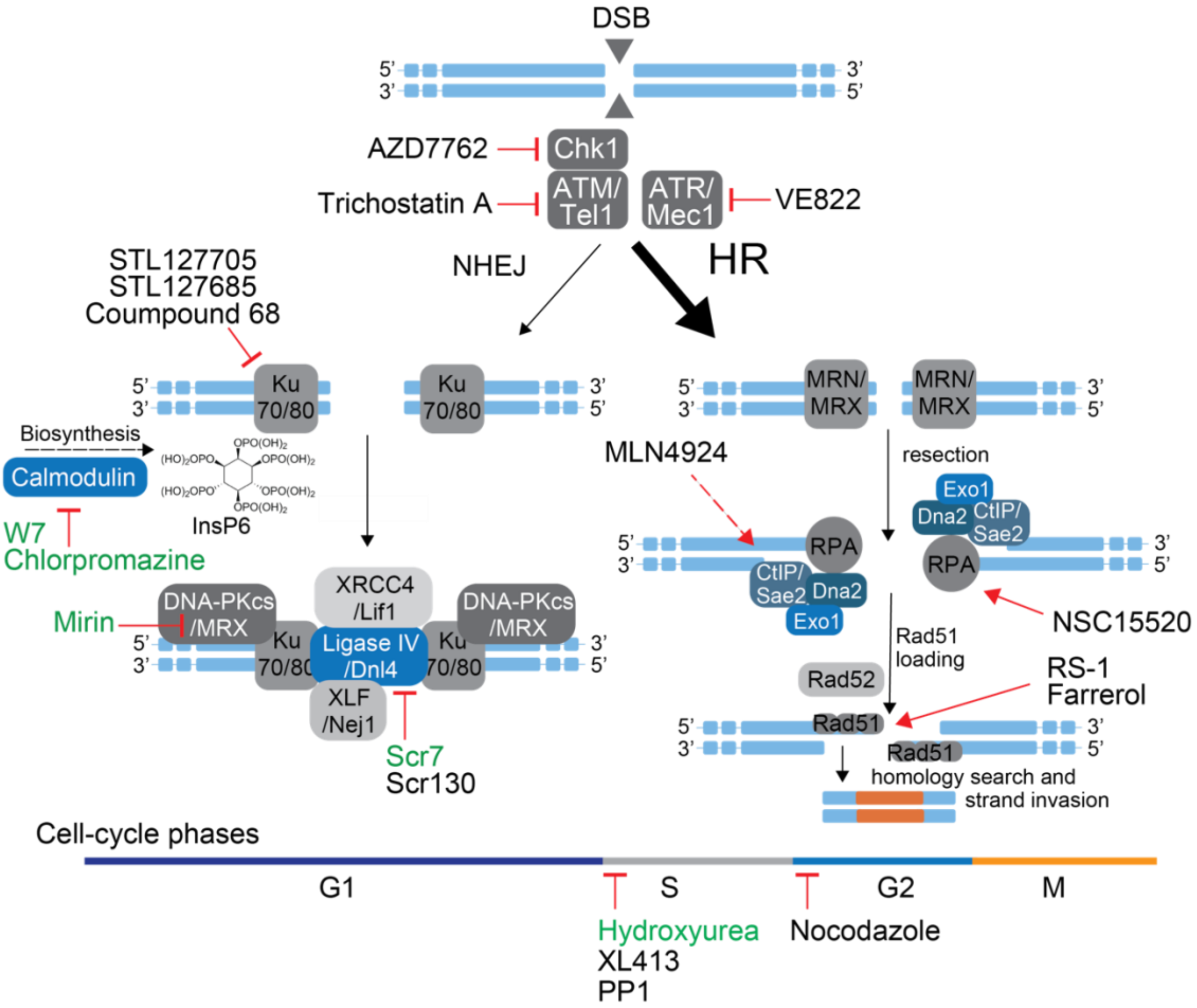

**Figure 2. Small molecules described or anticipated to target key proteins of NHEJ and HR in humans and in yeast as well as to interfere with cell-cycle progression.** The figure gives a simplified overview of the proteins involved in each pathway. For readability, some proteins and protein interactions are not shown. Small molecule inhibitors or activators that have been tested in yeast are shown in green. An overview of the molecules and their targets can also be found in **Table 1.** The molecule MLN4924 indirectly activates CtlP – indicated by a dashed arrow – by inhibiting the NEDD8-activating enzyme (NAE) and thus inhibiting the Neddylation of CtlP.

## 3. HR activators.

A detailed overview of homologous repair in yeast and humans can be found here (Li, et al., 2008). In short, after the detection of a double-strand break (DSB), the MRN/MRX complex is loaded onto the DNA ends, and subsequently several helicases and nucleases, including Mre11, Sae2, Exo1, Dna2, and Sgs1 resect the DNA ends to create 3′ overhangs **(Figure 2).** RPA-coated single-stranded DNA overhangs bind Rad52 and drive Rad51 filament formation. Rad51 catalyzes the key reactions that typify HR: homology search and DNA strand invasion. As Rad51 is a major player in HR, most current chemical strategies in mammalian cells target this protein. To the best of our knowledge, none of the HR enhancers has been tested in yeast.

**3.1. Stimulation of Rad51 enhances HR in mammalian cells but has not been tested in yeast.** Overexpression of the *S. cerevisiae* Rad51 gene was shown to enhance HDR in mammalian cells (Vispe, et al., 1998), and several small molecules that enhance its function have been reported.

RS-1 was discovered in an HTP chemical library screen to enhance human Rad51 ssDNA binding activity in a wide range of biochemical conditions (Jayathilaka, et al., 2008) and RS-1 subsequently showed to increase HR-mediated KI rates in various studies: 3 to 6-fold in HEK293A cells dependent on locus and transfection method (Pinder, et al., 2015), 5-fold in rabbit embryos (Song, et al., 2016) and also a significant increase in bovine embryos was observed (Lamas-Toranzo, et al., 2020).

Another RAD51 inhibitor is the molecule Farrerol, which was isolated in a screen assaying CRISPR-mediated HR rates exposed to a small molecule library from herbs used in traditional Chinese medicine. The authors show that Farrerol likely acts via accelerating the recruitment of Rad51 to broken DNA ends and enhanced the insertion rate 2 to 2.8-fold in various human and mouse cell lines, performing better than Scr7 and RS-1 which were tested in parallel (Zhang, et al., 2020).

**3.2. A small molecule that potentially enhances the availability of RPA minimally enhances HR in mammalian cells but has not been tested in yeast.** The replication protein A (RPA) is an essential player during HR by binding to single-stranded DNA and initiating DNA end resection, it accumulates during S and G2/M phases (Kibe, et al., 2007; Yamane, et al., 2013). During an HTP chemical screen for inhibitors of the replication protein A (RPA), the small molecule NSC15520 was identified (Glanzer, et al., 2011). NSC 15520 prevents the association of RPA with p53 and RAD9, possibly increasing the abundance of RPA available, which could favor HR. The molecule does not affect its ssDNA binding capacity. In analogy to the small molecule MLN4942 (neddylation inhibitor), NSC15520 enhanced precise genome editing only minimally by 1.3 to 1.4-fold, but it worked additively in combination with other small molecules (Riesenberg, et al., 2018) **(Point 5).**

## 4. Cell-cycle synchronizers

The fact that HR predominates in the S and the G2 phases of the cell cycle builds an opportunity to obtain a higher HR efficiency by reversibly arresting cells in one of the two phases.

**4.1 S-phase arrest via Hydroxyurea (HU) increases the efficiency of HDR-mediated genome editing in various yeast species but not in the one tested human cell line.** HU leads to S-phase arrest by inhibiting the enzyme ribonucleoside diphosphate reductase, thereby depleting cells of deoxyribonucleotides and thus limiting *de novo* DNA synthesis, leading to stalled replication forks and S-phase arrest (Madaan, et al., 2012). HU has successfully been used to increase homology-directed gene integration in yeast: The HR efficiency was increased in a species-dependent manner by 1.2- to 8-fold in *Arxula adeninivorans*, *S. cerevisiae*, *Kluveromyces lactis*, and *Pichia pastoris* even when using short homology arms (~50bp). In *Candida intermedia* integration rates were boosted from 1% to 70% using a split-marker genome editing approach (Peri, et al., 2023). In *Y. lipolytica* KI rates at 15 loci showed a large range of enhancements from 4 to 96% for different loci (Tsakraklides, et al., 2015). In a separate study in *Y. lipolytica*, HU increased integration rates to up to 100% using relatively short homology arms of 100bps, when additionally Ku70 was deleted (Jong, et al., 2018). However, only a 4% increase in HR frequency was observed in *M. pulcherrima* (Moreno-Beltran, et al., 2021). HU showed no improvement in HR-mediated genome engineering in HEK293T cells (Lin, et al., 2014).

**4.2 S-phase arrest via CDC7 inhibitors enhances genome integration in various mammalian cells but has not been tested in yeast.** Cell division cycle 7-related protein (Cdc7) kinase is a key regulator in the initiation of DNA replication and important for cell cycle progression into the G1 phase, inhibition of CDC7 leads to S-phase arrest (Masai, et al., 2002). Wienert *et al.* performed a screen using commercially available small molecule inhibitors of important cell-cycle proteins including CDC7 and found that XL413 – a known CDC7 inhibitor – increased CRISPR-mediated HR up to 3.5-fold across multiple loci and in multiple cell lines, such as primary T cells, Hela, and HSPCs (Wienert, et al., 2020). They compared efficiency with other inhibitors and in their work XL413 performed better than other molecules discussed herein such as Scr7 or RS1 (Wienert, et al., 2020). In an independent study, XL413 increased the HR-mediated gene insertion efficiency by 46% in primary human T cells (Shy, et al., 2022). Despite CDC7 being strongly conserved across eukaryotes (Masai, et al., 2002), XL413 has not been tested in yeast.

Further of note, the molecule PP1 was demonstrated to reversibly inhibit CDC7 activity in yeast and arrest cells in the G1 phase but has not been tested in genome engineering studies (Wan, et al., 2006).

**4.3 Other cell cycle synchronizers enhance efficiencies in human cells but have not been tested in yeast.** Lin *et al.* tested six known reversible chemical inhibitors synchronizing HEK293T cells at the G1, S, and M phases. They found aphidicolin (early S arrest) and nocodazole (late G1 arrest) to enhance integration rates from 9% to almost 30% (Lin, et al., 2014). Nocodazole had also been found to enhance HR-mediated gene integration 2-fold in HEK293T cells in an earlier non-CRISPR-based genome engineering study (Katada, et al., 2012), and later Eghbalsaied *et al.* showed that Nocodazole increased

HR-mediated integration rate by 21% in a murine embryonic fibroblast cell line (Eghbalsaied, et al., 2023). Nocodazole binds to beta-tubulin and disrupts microtubule assembly/disassembly dynamics, impairing the formation of the metaphase spindles during the cell division cycle. This prevents mitosis by inducing a G2/M-phase arrest (Jordan, et al., 1992). Nocodazole has not been explored in yeast.

## 5. Combinations of chemical inhibitors and activators.

In addition, the combination of several small molecules with different modes of action has been successfully explored in mammalian cells. Riesenberg *et al.* showed that the so-called CRISPY mix (a combination of the molecules NU7026, Trichostatin A, MLN4924, and NSC15520) had additive effects on precise editing efficiencies in different human cell lines (iPSCs and hESCs) and across three loci when using a double nicking strategy with Cas9n. In combination, the mix reached up to 50% precise nucleotide exchanges when compared to 10-30% for single molecules. The CRISPY mix thereby combines NHEJ inhibitors, HR activators, and NHEJ/HR regulators **(Figure 2)**: It inhibits DNA-PKcs with NU7026 (no homolog in yeast), enhances PKA with NSC15520, enhances Cpf1 with MLN4924 and inhibits ATM via Trichostatin A (Riesenberg, et al., 2018). Further, RS-1 was shown to work additive with Scr7, a combination that led to a 74% increase in HR efficiency (Zhang, et al., 2018).

**Table1. Small molecules described or anticipated to improve HR-mediated genome editing in mammalian cells and which have potential targets in yeast**. For simplicity, the *S. cerevisiae* gene name was chosen when indicating the yeast target. Based on the *S. cerevisiae* target, homologs in non-conventional yeast can be identified.

| Small molecules | Target | Target in yeast | Yeast species used for testing | Effect in HR in yeast | Reference |
|---|---|---|---|---|---|
| Inhibition of NHEJ | | | | | |
| STL127705 | Ku70/80 | yKu70/80 | - | - | (Weterings, et al., 2016; Guo, et al., 2022) |
| STL127685 | | | - | - | (Riesenberg, et al., 2018) |
| Compound 68 | | | - | - | (Gavande, et al., 2020; Mendoza-Munoz, et al., 2023) |
| W7 | Calmodulin/Ku-cofactor InsP6 | Calmodulin/InsP6 | *Cryptococcus neoformans*; *Metschnikowia pulcherrima* | 4-fold increase; 2% increase | (Arras, et al., 2016; Moreno-Beltran, et al., 2021) |
| Chlorpromazine | | | *C. neoformans* | 3- fold increase | |
| Scr7 | Ligase IV | Dnl4 | *C. neoformans* | - | (Maruyama, et al., 2015; Arras, et al., 2016; Song, et al., 2016; Gerlach, et al., 2018; Hu, et al., 2018; Zhang, et al., 2018; Anuchina, et al., 2023) |
| Scr130 | | | - | - | (Ray, et al., 2020) |
| Mirin | Mre11 | Mre11 | *C. neoformans* | 2-fold increase | (Arras, et al., 2016) |
| VE-822 | ATR | Mec1 | - | - | (Ma, et al., 2018) |
| AZD7762 | Chk1 | ScChk1 | - | - | (Ma, et al., 2018) |
| Trichostatin A | ATM | Tel1 | - | - | (Riesenberg, et al., 2018) |
| MLN4924 | NAE | Neddylation | - | - | (Riesenberg, et al., 2018) |
| Enhancement of Homologues Recombination | | | | | |
| RS-1 | Rad51 | Rad51 | - | - | (Pinder, et al., 2015; Song, et al., 2016; Lamas-Toranzo, et al., 2020) |
| Farrerol | Rad51 | Rad51 | - | - | (Zhang, et al., 2020) |
| NSC15520 | RPA70 | RPA | - | - | (Riesenberg, et al., 2018) |
| Cell-cycle synchronization | | | | | |
| Hydroxyurea | S phase | S phase | *Candida intermedia* | 70% efficiency with split-marker approach | (Peri, et al., 2023) |
| | | | *Arxula adeninivorans S. cerevisiae, Kluyveromyces lactis, Pichia pastori* | 1.2- to 8- fold increase | (Tsakraklides, et al., 2015) |
| | | | *Y. lipolytica* | 100% HR efficiency with disruption of Ku70 | (Jong, et al., 2018) |
| | | | *M. pulcherrima* | 4% increase | (Moreno-Beltran, et al., 2021) |
| XL413 | | | - | - | (Wienert, et al., 2020; Shy, et al., 2022) |
| PP1 | | | *S. cerevisiae SK1* | | (Wan, et al., 2006) |
| Nocodazole | G2 phase | G2 phase | - | - | (Katada, et al., 2012; Lin, et al., 2014; Eghbalsaied, et al., 2023) |

## 6. Assays that quantify frequencies of HR and NHEJ that could be employed for screening and the identification of yeast-specific chemicals.

Several assays have been developed to measure the frequency of NHEJ or HR events in mammalian or yeast cells that have been used to either screen for molecules that bias the repair (Arras, et al., 2016; Chen, et al., 2019; Ploessl, et al., 2022) or to better understand the molecular mechanisms behind it (Bindra, et al., 2013; Hussain, et al., 2021). We highlight a few that are medium to high throughput compatible and that could be repurposed to measure or screen the effect of chemical agents on repair mechanisms in yeast. Other versions of these assays which are not specifically discussed here can be found in (Pierce, et al., 1999; Pastwa, et al., 2009; Gunn, et al., 2012; Soong, et al., 2015; Wienert, et al., 2020).

**6.1. Screening for DSB repair inhibition by measuring synergy between chemical inhibitors and DNA damaging agents – no genetics required.** Small molecules that effectively inhibit DSB repair are toxic to cells at high concentrations (Arras, et al., 2016). As such, the minimal inhibitory concentration (MIC) of a given chemical should first be determined such that working concentrations with no growth effect can be chosen. Further, inhibition of NHEJ sensitizes cells towards DNA-damaging agents, as those are detrimental in case DSBs cannot be repaired. As such, Arras *et al.* (Arras, et al., 2016) used a simple assay measuring synergistic growth inhibition by a given small molecule in combination with a DSB-inducing agent as a first indication that a small molecule affects DSB repair. The authors specifically test phleomycin, mercaptopurine, and hydroxyurea as DNA-damaging agents in combination with various potential DSB repair inhibitors in the yeast *C. neoformans*. This assay is a

simple growth assay that can be performed in microtiter plates and does not require access to the genetics of the host.

### 6.2. Integration/deletion assays based on selectable phenotypes – no prior access to genetics required.

*ADE2* is a widely used non-selective marker gene as its deletion or the disruption of the encoding ORF leads to visibly pink cells when grown on media with low amounts of adenine. *ADE2* encodes for the enzyme phosphoribosylaminoimidazole carboxylase (Ade2) which catalyzes a step in the purine nucleotide biosynthesis, if dysfunctional, a visibly pink pigment accumulates in cells. In combination with an antibiotic resistance marker such as the gene coding for Nourseothricin N-acetyl transferase ($NAT^R$), it can be used as a screen to distinguish the repair of a DSB in the *ADE2* locus by NHEJ or HR **(Figure 3B).** For example, cells can be transformed with a linear PCR-generated DNA encoding for the *NAT* gene flanked by *ADE2-directed* homology arms. If the DSB is repaired by HR, *NAT* gets inserted into the *ADE2* locus and cells are both pink and resistant to nourseothricin, while NHEJ-repaired cells are just pink, in case the open reading frame is disrupted by Indels (Arras, et al., 2016).

Moreno-Beltrán *et al.,* used a similar approach, just using the *URA3* gene as a maker instead of *ADE2*. HR-repaired DSB yield cells are both, auxotrophic for uracil and nourseothricin resistant, while NHEJ-repaired cells are just auxotrophic for uracil in case the Ura3 open reading frame is disrupted (Moreno-Beltran, et al., 2021)

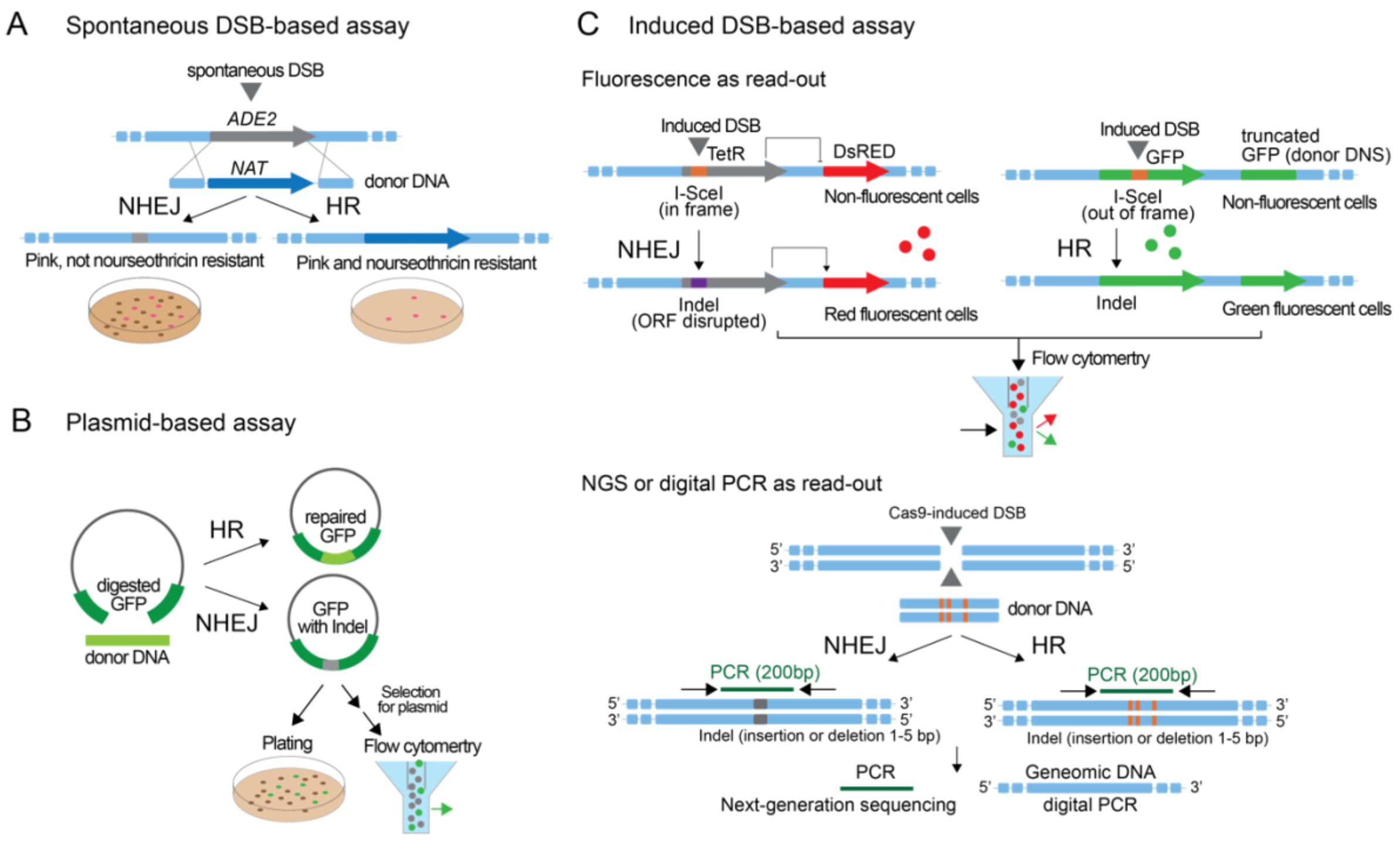

**Figure 3: Potential medium- and high-throughput (MTP/HTP) compatible assays to measure NHEJ/HR frequencies during small molecule screens in yeast. A.** *ADE2* encodes for phosphoribosylaminoimidazole carboxylase (Ade2) which catalyzes a step in the purine nucleotide biosynthesis. If deleted a pink pigment accumulates in cells on media with limiting amounts of supplemented adenine (amounts need to be titrated to receive a balance between cell growth and pigment production) that is visible to the naked eye and has been widely used as a read-out. After a spontaneous DSB, HR with a donor DNA that encodes for a Nourseothricin resistance marker (*NAT*) yields pink colonies that are also resistant to Nourseothricin. Repair by NHEJ yields just pink colonies as the Ade2 ORF can be disrupted to due Indels. This screen does not require *a priori* engineered yeast strains. **B.** The plasmid-based assay is based on the repair of a restriction-digested GFP that is encoded on a plasmid, with a donor DNA fragment during the co-transformation of a given yeast with both fragments. The donor DNA can be generated by PCR using primer-encoded homology arms of varying lengths. **C:** A more quantitative measurement of NHEJ vs. HDR after an induced double-strand break can be achieved via fluorescent devices and flow cytometry (upper panel) or NGS/digital PCR (lower panel). Note that in these systems the cells need to express a genome editing system such as the I-SceI endonuclease or a CRISPR/Cas9 system from a plasmid or the genome. **Upper panel:** The NHEJ reporter is based on a TetR repressor with an in-frame I-Sce-I site that represses a red fluorescent protein. In case the I-SecI site is cut, error-prone repair by NHEJ should yield a dysfunctional TetR and thus de-repression of the red fluorescent protein reporter, quantifiable by flow cytometry. The HR sensor is based on a mutated GFP that carries an out-of-frame I-SceI site to be repaired by a downstream encoded truncated GFP that serves as an intragenomic donor DNA giving rise to a functional GFP gene, and GFP fluorescence measurable by flow cytometry. **Lower panel:** After cells have undergone repair of the Cas9-generated break, the locus is PCR amplified around the break site (~200 bp product) and sequenced using Illumina-based NGS. Alternatively, extracted genomic DNA is used as an input for droplet digital PCR (ddPCR). Bioinformatic analysis allows quantification of each repair event: NHEJ is thereby classified as short deletions ranging from 1–5 bp at the break site, and HR is characterized as a specific replacement with the repair oligo and its 3 mutations.

**6.3. Plasmid-based assays to quantify HR and NHEJ – plasmid transformation required.** Ploessl *et al.* developed an assay able to detect if a cut plasmid gets repaired by NHEJ or HR (Ploessl, et al., 2022). It is based on transforming a plasmid where the encoded GFP has been cut with two restriction enzymes, leaving a linear plasmid with a truncated GFP. A donor DNA with 30 bp homology that encodes the missing piece is co-transformed. Repair via HR using the donor DNA leads to the repair of the GFP and consequently, green fluorescent transformants, while repair via NHEJ leads to simple plasmid circularization and non-fluorescent transformants **(Figure 3)**. The results can be quantified by fluorescent/non-fluorescent colony counting after plating on plasmid-selective media using a transilluminator or by quantifying fluorescent/non-fluorescent single cells using flow cytometry of plasmid selection in liquid media for several generations (higher throughput readout and less laborious). The assay could be performed in the presence or absence of various chemicals under investigation. The authors developed plasmids for transforming four yeast (*S. cerevisiae*, *K. marxianus*, *S. stipitis*, *Y. lipolytica*). The systems could likely be extended to other yeast given transformation protocols, a selection marker, and a strong promoter driving GFP expression are available. The authors assume that the assay underrepresents preference for NHEJ in a given yeast in comparison to repairing a genomic DSB e.g. induced by Cas9 as genomic DNA is a more complex substrate. Further of note, the authors use rather short homology arms (30bp) in comparison to other studies and the sensitivity of the assay could potentially be enhanced by longer homology arms.

**6.4. Genomic DSB-based assays for NHEJ and HR quantification – prior plasmid transformation and/or genome engineering required.** We highlight examples that allow quantifying NHEJ and HR after an induced DSB on the genome; they either use fluorescent proteins and flow cytometry for quantification or next-generation sequencing (NGS) and droplet digital PCR (ddPCR) for quantification, the latter is cheaper and less bioinformatically demanding then NGS **(Figure 3C).**
Bindra *et al.* developed a fluorescent-based assay (called EJ-RFP) to detect NHEJ and verify it across various human cell lines (Bindra, et al., 2013). It relies on inducing a genomic DSB by the rare-cutting endonuclease I-SceI (should be replaceable by a Cas-system) and a genomically integrated reporter construct: A TetR repressor with an in-frame I-Sce-I site that represses the expression of a red fluorescent protein. In case the I-SecI site is cut, error-prone repair by NHEJ should yield a dysfunctional TetR and thus de-repression of the red fluorescent protein reporter, quantifiable by flow cytometry **(Figure 3C, upper panel)**. The authors combine their NHEJ-detecting EJ-RFP assay with a previously developed HR-detecting "direct repeat green fluorescent protein" (DR-GFP) assay to quantify the ratio of both events in cells. In the DR-GFP assay, the I-SceI site has been integrated into a GFP gene thereby disrupting its open reading frame (ORF). A truncated GFP fragment with the correct ORF sequence has been placed downstream in the construct and serves as an intragenomic donor DNA. Repair of the cleaved I-SceI site by HR using the downstream fragment gives rise to a functional GFP gene, and GFP fluorescence then can be measured by flow cytometry.

Hussain *et al.* (Hussain, et al., 2021) developed a Cas9-based system that quantifies repair outcomes of NHEJ, HR, and a third alternative "backup pathway", alternative end-joining (Alt-EJ), which has not been further discussed here. For the assay, cells need to be equipped with a Cas9/gRNA encoding system either on a plasmid or genomically integrated. Cells are then co-transformed with a single-stranded and a double-stranded ~200 bp repair oligonucleotide with three mutations. After cells have undergone repair of the Cas9-generated break, the locus is PCR amplified and sequenced using Illumina-based NGS. Alternatively, extracted genomic DNA is used as an input for droplet digital PCR (ddPCR). Bioinformatic analysis allows quantification of each repair event **(Figure 3C, lower panel).** This assay is HTP-compatible as many chemicals could be tested in parallel, it can be used to test repair at any locus that can be targeted by Cas9 and does not require genomically integrated reporter cassettes. The authors develop the assay in human cell lines, but it should also be applicable to yeast.

## 7. Discussion

The use of small molecules has proven promising for achieving higher success rates in the homology-directed engineering of human cells without the need to modify the genome of the host. Not only human engineering suffers from the fact that NHEJ dominates over HR but it also impairs the implementation of modern genome engineering tools in many unconventional yeasts. Given the conserved proteins in repair pathways in mammal cells and yeast cells, researchers started exploring these mammalian-optimized compounds for yeast engineering. Even though not widely used yet, a few successful stories have been reported **(Table 1).**

The compound that has shown the most consistent success across yeast species and various studies is the cell-cycle synchronizer hydroxyurea which arrests cells in the S phase of the cell cycle where HR dominates over NHEJ. But also, the calmodulin inhibitors W7 and Chlorpromazine, the ligase IV inhibitor Src7, and the Mre11 inhibitor Mirin showed promising results in yeast. Many compounds that showed success in humans have not yet been explored in yeast **(Table 1).** In addition, the combination of several small molecules with different modes of action could be explored, such as successfully shown for mammalian cells (Riesenberg, et al., 2018;  Zhang, et al., 2018).

Importantly, in the instances where small molecules have been used in yeast to raise HR efficiency, the authors have not yet combined them with CRISPR/Cas systems but rather used pure homolog recombination-based techniques in combination with rather long homology arms and selection markers. As such, combination with state-of-the-art genome editing tools, especially those recently developed for unconventional yeast (Ploessl, et al., 2022) could be very promising to generate broad-species protocols that enhance HR. Also, natural strain-to-strain variability in HR efficiency needs to be considered when starting to engineer a new yeast species (e.g. testable via the assay outlined in **Figure 3A**). An isolate screen in *M. pulcherrima* exemplifies that only a few tested isolates were amenable to HR-mediated gene insertions and enhancement via chemical inhibitors (Moreno-Beltran, et al., 2021).

Currently, available small molecules have been optimized for human cells. They have been either derived by HTP screens using available small molecule libraries and various purpose-directed read-out assays or via molecular modeling/docking using available 3D structures of human protein targets or available homologs. Both strategies could be employed to derive yeast-specific molecules. We summarized several NHEJ/HR detection assays that could be employed in small-molecule screens, as well as structure prediction tools that could be used to gain access to 3D structures of repair proteins of various yeast species for docking and chemical refinement studies.

One aspect is still important to discuss: It cannot be ignored that few of the small molecules employed in humans and yeast have been shown to increase HR efficiency in all cell types or yeast species. In fact, some small molecules can have non-identical and even opposite effects on precise genome-editing efficiencies across cell types. The reported efficiencies of the various chemicals in human cells are difficult to compare given the diversity in used cell lines, loci, small molecule concentrations, and protocols. As such, for yeast engineering, it would be meaningful to use standardized measurements and assays to measure HR-related genome engineering, to gain lab-to-lab reusable protocols, and to better understand variability in success rates associated with a certain chemical agent.

Further, the potential toxicity of a given small molecule needs to be considered. For example, Song *et al.*, found severe cell death in human induced pluripotent stem cells when treated with compound Scr7 (Song, et al., 2016), while no adverse effects were observed in mouse embryos and several human cancer cell lines (Singh, et al., 2015;  Hu, et al., 2018). Toxicity may be related to the concentration and to the specific cell type and as such the minimal inhibitory concentration should be determined before use in a new yeast, as exemplified for *C. neoformans* (Arras, et al., 2016). Besides that, some molecules show low solubility in DMSO as such compatible solvents need to be screened.

## Reference

1. Anuchina, A. A., Zaynitdinova, M. I., Demchenko, A. G., Evtushenko, N. A., Lavrov, A. V., and Smirnikhina, S. A. (2023) Bridging gaps in hdr improvement: The role of MAD2L2, SCAI, and Scr7. Int J Mol Sci 24.
2. Arras, S. D. M., and Fraser, J. A. (2016) Chemical inhibitors of non-homologous end joining increase targeted construct integration in *Cryptococcus neoformans*. PloS One 11.
3. Azhagiri, M. K. K., Babu, P., Venkatesan, V., and Thangavel, S. (2021) Homology-directed gene-editing approaches for hematopoietic stem and progenitor cell gene therapy. Stem Cell Res Ther 12.
4. Besada-Lombana, P. B., McTaggart, T. L., and Da Silva, N. A. (2018) Molecular tools for pathway engineering in *Saccharomyces cerevisiae*. Curr Opin Biotechnol 53: 39-49.
5. Billerbeck, S., Prins, R. C., and Marquardt, M. (2023) A modular cloning toolkit including CRISPRi for the engineering of the human fungal pathogen and biotechnology host *Candida glabrata*. Acs Synth Biol 12: 1358-1363.
6. Bindra, R. S., Goglia, A. G., Jasin, M., and Powell, S. N. (2013) Development of an assay to measure mutagenic non-homologous end-joining repair activity in mammalian cells. Nucleic Acids Res 41: e115.
7. Brachmann, C. B., Davies, A., Cost, G. J., Caputo, E., Li, J. C., Hieter, P., and Boeke, J. D. (1998) Designer deletion strains derived from *Saccharomyces cerevisiae* S288C: a useful set of strains and plasmids for pcr-mediated gene disruption and other applications. Yeast 14: 115-132.
8. Byrum, J., Jordan, S., Safrany, S. T., and Rodgers, W. (2004) Visualization of inositol phosphate-dependent mobility of Ku: depletion of the DNA-PK cofactor insp(6) inhibits Ku mobility. Nucleic Acids Res 32: 2776-2784.
9. Cai, P., Gao, J. Q., and Zhou, Y. J. (2019) CRISPR-mediated genome editing in non-conventional yeasts for biotechnological applications. Microb Cell Factories 18.
10. Cannan, W. J., and Pederson, D. S. (2016) Mechanisms and consequences of double-strand DNA break formation in chromatin. J Cell Physiol 231: 3-14.
11. Chen, S. W., Chen, D., Liu, B., and Haisma, H. J. (2022) Modulating CRISPR/Cas9 genome-editing activity by small molecules. Drug Discov Today 27: 951-966.
12. Chen, Y., Zhang, H. P., Xu, Z., Tang, H. Y., Geng, A. K., Cai, B. L., et al. (2019) A PARP1-BRG1-SIRT1 axis promotes HR repair by reducing nucleosome density at DNA damage sites. Nucleic Acids Res 47: 8563-8580.
13. Davis, T. N., and Thorner, J. (1989) Vertebrate and yeast calmodulin, despite significant sequence divergence, are functionally interchangeable. Proc Natl Acad Sci USA 86: 7909-7913.
14. Denobel, J. G., and Barnett, J. A. (1991) Passage of molecules through yeast-cell walls - a brief essay-review. Yeast 7: 313-323.

15. Dudášová, Z., Dudáš, A., and Chovanec, M. (2004) Non-homologous end-joining factors of *Saccharomyces cerevisiae*. FEMS Microbiol Rev 28: 581-601.

16. Dupre, A., Boyer-Chatenet, L., Sattler, R. M., Modi, A. P., Lee, J. H., Nicolette, M. L., et al. (2008) A forward chemical genetic screen reveals an inhibitor of the MRE11-RAD50-NBS1 complex. Nat Chem Biol 4: 119-125.

17. Eghbalsaied, Shahin and Kues, Wilfried A. (2023) CRISPR/Cas9-mediated targeted knock-in of large constructs using nocodazole and RNase HII. Sci Rep 13: 2690.

18. Fisher, M. C., Gow, N. A. R., and Gurr, S. J. (2016) Tackling emerging fungal threats to animal health, food security and ecosystem resilience. Philos Trans R Soc Lond B Biol Sci 371.

19. Gabaldon, T., Arastehfar, A., Boekhout, T., Butler, G., De Cesare, G. B., Dolk, E., et al. (2019) Recent trends in molecular diagnostics of yeast infections: from PCR to NGS. FEMS Microbiol Rev 43: 517-547.

20. Gavande, N. S., VanderVere-Carozza, P. S., Pawelczak, K. S., Mendoza-Munoz, P., Vernon, T. L., Hanakahi, L. A., et al. (2020) Discovery and development of novel DNA-PK inhibitors by targeting the unique Ku-DNA interaction. Nucleic Acids Research 48: 11536-11550.

21. Gerlach, M., Kraft, T., Brenner, B., Petersen, B., Niemann, H., and Montag, J. (2018) Efficient knock-in of a point mutation in porcine fibroblasts using the CRISPR/Cas9-GMNN fusion gene. Genes 9.

22. Gibson, D. G., Benders, G. A., Axelrod, K. C., Zaveri, J., Algire, M. A., Moodie, M., et al. (2008) One-step assembly in yeast of 25 overlapping DNA fragments to form a complete synthetic *Mycoplasma genitalium* genome. Proc Natl Acad Sci USA 105: 20404-20409.

23. Glanzer, J. G., Liu, S. Q., and Oakley, G. G. (2011) Small molecule inhibitor of the RPA70 N-terminal protein interaction domain discovered using in silico and in vitro methods. Bioorg Med Chem 19: 2589-2595.

24. Greco, G. E., Matsumoto, Y., Brooks, R. C., Lu, Z. F., Lieber, M. R., and Tomkinson, A. E. (2016) Scr7 is neither a selective nor a potent inhibitor of human DNA ligase IV. DNA Repair 43: 18-23.

25. Guo, N. N., Li, S. J., Liu, B., Chen, P., Li, J. B., Zhao, Y. L., et al. (2022) Inhibiting nonhomologous end-joining repair would promote the antitumor activity of gemcitabine in nonsmall cell lung cancer cell lines. Anti-Cancer Drugs 33: 502-508.

26. Hengel, S. R., Spies, M. A., and Spies, M. (2017) Small-molecule inhibitors targeting DNA repair and DNA repair deficiency in research and cancer therapy. Cell Chem Biol 24: 1101-1119.

27. Hu, Z., Shi, Z. Y., Guo, X. G., Jiang, B. S., Wang, G., Luo, D. X., et al. (2018) Ligase IV inhibitor Scr7 enhances gene editing directed by CRISPR-Cas9 and ssODN in human cancer cells. Cell Biosci 8.

28. Hussain, S. S., Majumdar, R., Moore, G. M., Narang, H., Buechelmaier, E. S., Bazil, M. J., et al. (2021) Measuring nonhomologous end-joining, homologous recombination and alternative end-joining simultaneously at an endogenous locus in any transfectable human cell. Nucleic Acids Res 49: e74.

29. Jayathilaka, K., Sheridan, S. D., Bold, T. D., Bochenska, K., Logan, H. L., Weichselbaum, R. R., et al. (2008) A chemical compound that stimulates the human homologous recombination protein RAD51. Proc Natl Acad Sci USA105: 15848-15853.

30. Ji, Q. C., Mai, J., Ding, Y., Wei, Y. J., Ledesma-Amaro, R., and Ji, X. J. (2020) Improving the homologous recombination efficiency of *Yarrowia lipolytica* by grafting heterologous component from *Saccharomyces cerevisiae*. Metab Eng Commun 11.

31. Jin, M. H., and Oha, D. Y. (2019) ATM in DNA repair in cancer. Pharmacol Ther 203.

Jong, I. S., Yu, B. J., Jang, J. Y., Jegal, J., and Lee, J. Y. (2018) Improving the efficiency of homologous recombination by chemical and biological approaches in *Yarrowia lipolytica*. PloS One 13.

32. Jordan, M. A., Thrower, D., and Wilson, L. (1992) Effects of vinblastine, podophyllotoxin and nocodazole on mitotic spindles - implications for the role of microtubule dynamics in mitosis. J Cell Sci 102: 401-416.

33. Katada, Hitoshi, Harumoto, Toshimasa, Shigi, Narumi and Komiyama, Makoto. (2012) Chemical and biological approaches to improve the efficiency of homologous recombination in human cells mediated by artificial restriction DNA cutter. Nucleic Acids Res 40: e81.

34. Kavscek, M., Strazar, M., Curk, T., Natter, K., and Petrovic, U. (2015) Yeast as a cell factory: Current state and perspectives. Microb Cell Factories 14.

35. Kibe, T., Ono, Y., Sato, K., and Ueno, M. (2007) Fission yeast Taz1 and RPA are synergistically required to prevent rapid telomere loss. Mol Biol Cell 18: 2378-2387.

36. Kooistra, R., Hooykaas, P. J. J., and Steensma, H. Y. (2004) Efficient gene targeting in *Kluyveromyces lactis*. Yeast 21: 781-792.

37. Lamas-Toranzo, I., Martinez-Moro, A., O'Callaghan, E., Milian-Blanca, G., Sanchez, J. M., Lonergan, P., and Bermejo-Alvarez, P. (2020) RS-1 enhances CRISPR-mediated targeted knock-in in bovine embryost. Mol Reprod Dev 87: 542-549.

38. Lee, M. E., DeLoache, W. C., Cervantes, B., and Dueber, J. E. (2015) A highly characterized yeast toolkit for modular, multipart assembly. Acs Synth Biol 4: 975-986.

39. Li, X., and Heyer, W. D. (2008) Homologous recombination in DNA repair and DNA damage tolerance. Cell Res 18: 99-113.

40. Lin, S., Staahl, B., Alla, R. K., and Doudna, J. A. (2014) Enhanced homology-directed human genome engineering by controlled timing of CRISPR/Cas9 delivery. Elife 3.

41. Ma, X. J., Chen, X., Jin, Y., Ge, W. Y., Wang, W. Y., Kong, L. H., et al. (2018) Small molecules promote CRISPR-Cpf1-mediated genome editing in human pluripotent stem cells. Nat Commun 9.

42. Madaan, K., Kaushik, D., and Verma, T. (2012) Hydroxyurea: a key player in cancer chemotherapy. Expert Rev Anticancer Ther 12: 19-29.

43. Maghames, C. M., Lobato-Gil, S., Perrin, A., Trauchessec, H., Rodriguez, M. S., Urbach, S., et al. (2018) Neddylation promotes nuclear protein aggregation and protects the ubiquitin proteasome system upon proteotoxic stress. Nat Commun 9.

44. Malci, K., Walls, L. E., and Rios-Solis, L. (2020) Multiplex genome engineering methods for yeast cell factory development. Front Bioeng Biotechnol 8.

45. Maruyama, T., Dougan, S. K., Truttmann, M. C., Bilate, A. M., Ingram, J. R., and Ploegh, H. L. (2015) Increasing the efficiency of precise genome editing with CRISPR-Cas9 by inhibition of nonhomologous end joining. Nat Biotechn 33: 538-542.

46. Masai, H., and Arai, K. I. (2002) Cdc7 kinase complex: a key regulator in the initiation of DNA replication. J Cell Physiol 190: 287-296.

47. Mendoza-Munoz, P. L., Gavande, N. S., VanderVere-Carozza, P. S., Pawelczak, K. S., Dynlacht, J. R.,

48. Garrett, J. E. and Turchi, J. J. (2023) Ku-DNA binding inhibitors modulate the DNA damage response in response to DNA double-strand breaks. NAR Cancer 5:

49. Moreno-Beltran, M., Gore-Lloyd, D., Chuck, C., and Henk, D. (2021) Variation among *Metschnikowia pulcherrima* isolates for genetic modification and homologous recombination. Microorganisms 9.

50. Nambiar, T. S., Billon, P., Diedenhofen, G., Hayward, S. B., Taglialatela, A., Cai, K. H., et al. (2019) Stimulation of CRISPR-mediated homology-directed repair by an engineered RAD18 variant. Nat Commun 10.

51. Nayak, T., Szewczyk, E., Oakley, C. E., Osmani, A., Ukil, L., Murray, S. L., et al. (2006) A versatile and efficient gene-targeting system for *Aspergillus nidulans*. Genetics 172: 1557-1566.

52. Nielsen, J. (2019) Yeast systems biology: model organism and cell factory. Biotechnol J 14.

Ninomiya, Y., Suzuki, K., Ishii, C., and Inoue, H. (2004) Highly efficient gene replacements in neurospora strains deficient for nonhomologous end-joining. Proc Natl Acad Sci USA 101: 12248-12253.

53. Park, J., Kim, I. J., and Kim, S. R. (2022) Nonconventional yeasts engineered using the CRISPR-Cas system as emerging microbial cell factories. Fermentation-Basel 8.

54. Pastwa, E., Somiari, R. I., Malinowski, M., Somiari, S. B., and Winters, T. A. (2009) In vitro non-homologous DNA end joining assays-the 20th anniversary. Int J Biochem Cell Biol 41: 1254-1260.

55. Peri, K. V. R., Faria-Oliveira, F., Larsson, A., Plovie, A., Papon, N., and Geijer, C. (2023) Split-marker-mediated genome editing improves homologous recombination frequency in the CTG clade yeast *Candida intermedia*. FEMS Yeast Res 23.

56. Pierce, A. J., Johnson, R. D., Thompson, L. H., and Jasin, M. (1999) XRCC3 promotes homology-directed repair of DNA damage in mammalian cells. Genes Dev 13: 2633-2638.

57. Pinder, J., Salsman, J., and Dellaire, G. (2015) Nuclear domain 'knock-in' screen for the evaluation and identification of small molecule enhancers of CRISPR-based genome editing. Nucleic Acids Res 43: 9379-9392.

58. Ploessl, D., Zhao, Y. X., Cao, M. F., Ghosh, S., Lopez, C., Sayadi, M., et al. (2022) A repackaged CRISPR platform increases homology-directed repair for yeast engineering. Nat Chem Biol 18: 38-46.

59. Ray, U., Raul, S. K., Gopinatha, V. K., Ghosh, D., Rangappa, K. S., Mantelingu, K., and Raghavan, S. C. (2020) Identification and characterization of novel Scr7-based small-molecule inhibitor of DNA end-joining, Scr130 and its relevance in cancer therapeutics. Mol Carcinog 59: 618-628.

60. Ray, U., Vartak, S. V., and Raghavan, S. C. (2020) NHEJ inhibitor Scr7 and its different forms: promising CRISPR tools for genome engineering. Gene 763.

61. Riesenberg, S., and Maricic, T. (2018) Targeting repair pathways with small molecules increases precise genome editing in pluripotent stem cells. Nat Commun 9.

62. Saha, J., Wang, S. Y., and Davis, A. J. (2017) Examining DNA double-strand break repair in a cell cycle-dependent manner. DNA Repair Enzymes: Cell Mol Chem Biol 591: 97-118.

63. Schwarzmuller, T., Ma, B., Hiller, E., Istel, F., Tscherner, M., Brunke, S., et al. (2014) Systematic phenotyping of a large-scale *Candida glabrata* deletion collection reveals novel antifungal tolerance genes. PLoS Pathog 10.

64. Shams, F., Bayat, H., Mohammadian, O., Mahboudi, S., Vahidnezhad, H., Soosanabadi, M., and Rahimpour, A. (2022) Advance trends in targeting homology- directed repair for accurate gene editing: an inclusive review of small molecules and modified CRISPR- Cas9 systems. Bioimpacts 12: 371-391.

65. Shao, S.M., Ren, C. H., Liu, Z.T., Bai, Y.C., Chen, Z.L., Wei, Z.H., et al. (2017) Enhancing CRISPR/Cas9-mediated homology-directed repair in mammalian cells by expressing *Saccharomyces cerevisiae* RAD52. Int J Biochem Cell Biol 92: 43-52.

66. Shibata, A., Moiani, D., Arvai, A. S., Perry, J., Harding, S. M., Genois, M. M., et al. (2014) DNA double-strand break repair pathway choice is directed by distinct MRE11 nuclease activities. Mol Cell 53: 7-18.

67. Shy, B. R., Vykunta, V. S., Ha, A., Talbot, A., Roth, T. L., Nguyen, D. N., et al. (2022) High-yield genome engineering in primary cells using a hybrid ssDNA repair template and small-molecule cocktails. Nat Biotech 41: 521–531.

68. Singh, P., Schimenti, J. C., and Bolcun-Filas, E. (2015) A mouse geneticist's practical guide to CRISPR applications. Genetics 199: 1-15.

69. Song, J., Yang, D. S., Xu, J., Zhu, T. Q., Chen, Y. E., and Zhang, J. F. (2016) RS-1 enhances CRISPR/Cas9-and TALEN-mediated knock-in efficiency. Nat Commun 7.

70. Soong, C. P., Breuer, G. A., Hannon, R. A., Kim, S. D., Salem, A. F., Wang, G. L., et al. (2015) Development of a novel method to create double-strand break repair fingerprints using next-generation sequencing. DNA Repair 26: 44-53.

71. Soucy, T. A., Smith, P. G., Milhollen, M. A., Berger, A. J., Gavin, J. M., Adhikari, S., et al. (2009) An inhibitor of nedd8-activating enzyme as a new approach to treat cancer. Nature 458: 732-736.

72. Srivastava, M., Nambiar, M., Sharma, S., Karki, S. S., Goldsmith, G., Hegde, M., et al. (2012) An inhibitor of nonhomologous end-joining abrogates double-strand break repair and impedes cancer progression. Cell 151: 1474-1487.

73. Sui, J. D., Zhang, S. C., and Chen, B. P. C. (2020) DNA-dependent protein kinase in telomere maintenance and protection. Cell Mol Biol Lett 25.

74. Taheri-Ghahfarokhi, A., Taylor, B. J. M., Nitsch, R., Lundin, A., Cavallo, A. L., Madeyski-Bengtson, K., et al. (2018) Decoding non-random mutational signatures at Cas9 targeted sites. Nucleic Acids Res 46: 8417-8434.

75. Tsakraklides, V., Brevnova, E., Stephanopoulos, G., and Shaw, A. J. (2015) Improved gene targeting through cell cycle synchronization. PloS One 10.

76. Vispe, S., Cazaux, C., Lesca, C., and Defais, M. (1998) Overexpression of RAD51 protein stimulates homologous recombination and increases resistance of mammalian cells to ionizing radiation. Nucleic Acids Res 26: 2859-2864.

77. Wan, L. H., Zhang, C., Shokat, K. M., and Hollingsworth, N. M. (2006) Chemical inactivation of Cdc7 kinase in budding yeast results in a reversible arrest that allows efficient cell synchronization prior to meiotic recombination. Genetics 174: 1767-1774.

78. Wang, X. W., and Zhao, J. G. (2021) Targeted cancer therapy based on acetylation and deacetylation of key proteins involved in double-strand break repair. Cancer Manag Res 14: 259-271.

79. Weterings, E., Gallegos, A. C., Dominick, L. N., Cooke, L. S., Bartels, T. N., Vagner, J., et al. (2016) A novel small molecule inhibitor of the DNA repair protein Ku70/80. DNA Repair 43: 98-106.

80. Wienert, B., Nguyen, D. N., Guenther, A., Feng, S. J., Locke, M. N., Wyman, S. K., et al. (2020) Timed inhibition of Cdc7 increases CRISPR-Cas9 mediated templated repair. Nat Commun 11.

81. Yamane, A., Robbiani, D. F., Resch, W., Bothmer, A., Nakahashi, H., Oliveira, T., et al. (2013) RPA accumulation during class switch recombination represents 5 '-3 ' DNA-end resection during the S-G2/M phase of the cell cycle. Cell Rep 3: 138-147.

82. Zhang, W. N., Chen, Y., Yang, J. Q., Zhang, J., Yu, J. Y., Wang, M. T., et al. (2020) A high-throughput small molecule screen identifies farrerol as a potentiator of CRISPR/Cas9-mediated genome editing. Elife 9.

83. Zhang, Y. B., Zhang, Z. W., and Ge, W. (2018) An efficient platform for generating somatic point mutations with germline transmission in the zebrafish by CRISPR/Cas9-mediated gene editing. J Biol Chem 293: 6611-6622.